\documentstyle[12pt,epsfig]{article}
\catcode`\@=11
\def\numberbysection{\@addtoreset{equation}{section}
\def\theequation{\thesection.\arabic{equation}}}
\numberbysection

\newcommand{\abs}[1]{\left\vert#1\right\vert}
\newcommand{\be}{\[}
\newcommand{\ee}{\]}
\newcommand{\beq}{\begin{equation}}
\newcommand{\eeq}{\end{equation}}
\renewcommand{\d}{{\rm d}}
\newcommand{\e}{{\rm e}}
\newcommand{\esc}{{\rm esc}}
\renewcommand{\i}{{\rm i}}
\renewcommand\Im{\mathop{\rm Im}}
\newcommand\Frac{\mathop{\rm Frac}}
\newcommand{\F}{{\cal F}}
\newcommand{\G}{{\cal G}}
\renewcommand{\L}{{\rm L}}
\newcommand{\M}{{\cal M}}
\newcommand{\N}{{\cal N}}
\renewcommand{\O}{{\cal O}}
\newcommand{\bin}[2]{{#1\choose#2}}
\newcommand{\ad}{\normalsize\rm}

\title{\LARGE\bf Even-visiting random walks:
exact and asymptotic results in one dimension}

\author{M.~Bauer, D.~Bernard\footnote{Member of C.N.R.S.}, and J.M.~Luck
\bigskip
\\ \ad Service de Physique Th\'eorique, CEA Saclay,
91191 Gif-sur-Yvette cedex, France
\\ \ad email: bauer, dbernard, luck@spht.saclay.cea.fr}

\date{ }
\begin{document}
\maketitle
\begin{abstract}
We reconsider the problem of even-visiting random walks in one dimension.
This problem is mapped onto a non-Hermitian Anderson model
with binary disorder.
We develop very efficient numerical tools to enumerate
and characterize even-visiting walks.
The number of closed walks is obtained as an exact integer
up to 1828 steps, i.e., some $10^{535}$ walks.
On the analytical side, the concepts and techniques
of one-dimensional disordered systems allow to obtain explicit
asymptotic estimates for the number of closed walks of $4k$ steps
up to an absolute prefactor of order unity, which is determined numerically.
All the cumulants of the maximum height reached by such walks
are shown to grow as $k^{1/3}$, with exactly known prefactors.
These results illustrate the tight relationship
between even-visiting walks, trapping models,
and the Lifshitz tails of disordered electron or phonon spectra.
\end{abstract}
\vfill
\noindent To be submitted for publication to Journal of Physics A
\hfill T/01/001

\noindent P.A.C.S.: 05.40.Fb, 02.50.-r
\newpage
\section{Introduction}

Random walk is one of the most ubiquitous concepts of statistical physics.
The large-time behavior of random walks is rather insensitive to details
such as the presence of an underlying lattice
or of an elementary time step, so that a vast range of them belongs
to the universality class of Brownian motion,
their continuum limit being described by the diffusion equation.

Recently, in the context of surface growth, Noh, Park,
and den Nijs~\cite{dennijs} have introduced a model of random walks with the
constraint that every visited site should be visited an even number of times.
Such walks have been named even-visiting walks~\cite{ital1,ital2}.
It has been argued that the non-local constraint of being even-visiting
changes the universality class of random walks,
with the typical extension of a walk of $n$ steps scaling as
$n^{1/(d+2)}$ in dimension $d$, instead of the usual $n^{1/2}$.

Derrida's argument~\cite{derrida} for this behavior,
to be recalled just below,
demonstrates a deep analogy between the even-visiting walk problem,
the trapping problem~\cite{hk,nb}
of a diffusive particle which is absorbed by traps located at random positions,
and the Lifshitz tails~\cite{lif}
of the density of states of electron or phonon spectra
of disordered solids near band edges.

Consider for definiteness a random walk of $n$ steps
on the hypercubic lattice in dimension $d$, with unit lattice spacing.
The probability that this walk is confined within a sphere
with radius $R\ll n^{1/2}$ scales as
\be
p(R)\sim\exp\left(-\frac{j^2\,n}{2d\,R^2}\right),
\ee
where $j$ is the first positive zero of the Bessel function $J_{(d-2)/2}$,
so that $j^2/R^2$ is the lowest eigenvalue of the Laplace operator
in the sphere, with Dirichlet boundary conditions.
Then, if $n\gg R^d$, most sites in the sphere
must have been visited many times.
Hence, to a good approximation,
the probability that any given site has been visited an
even number of times is $1/2$, independently of other sites.
Thus, among all walks that remain within the sphere,
the fraction of even-visiting ones is about
\be
f(R)\sim\exp\left(-\Omega\ln 2\,R^d\right),
\ee
where $\Omega=\pi^{d/2}/\Gamma(d/2+1)$ is the volume of the unit ball.
Consequently, for large~$n$,
the main contribution to the number of even-visiting walks
comes from `Lifshitz spheres', i.e., optimal spheres
whose radius $R_\L$ maximizes the product $p(R)f(R)$, hence
\beq
R_\L\approx\left(\frac{j^2\,n}{d^2\,\Omega\ln 2}\right)^{1/(d+2)}.
\label{jr}
\eeq
Both inequalities needed to justify the argument are satisfied
by this solution, in any dimension $d$.
Eq.~(\ref{jr}) gives an estimate for the typical maximum extent
of an $n$-step even-visiting walk from its origin.
The total number of such walks scales as $(2d)^np(R_\L)f(R_\L)$, hence
\beq
\N(n)\sim(2d)^n\exp\left(-\frac{d+2}{2}(\Omega\ln 2)^{2/(d+2)}
\left(\frac{j^2\,n}{d^2}\right)^{d/(d+2)}\right).
\label{jn}
\eeq

In this paper we reconsider the one-dimensional case~\cite{dennijs,ital1,ital2}
in detail.
We map the even-visiting walk problem onto a model with quenched disorder,
analogous to a non-Hermitian Anderson model.
Many concepts and techniques
of one-dimensional disordered systems are then available~\cite{bl,cpv,jm}.
We develop efficient numerical tools to enumerate even-visiting walks.
The parameter in the generating function of the walks governs the strength of
disorder.
The disordered system comes with an invariant measure,
whose support is bounded for small disorder,
and becomes unbounded at some critical point,
at which the generating function of even-visiting walks develops a singularity.
We first compute the escape probability~(\ref{escap}),
which is similar to the integrated density of states of the Anderson model.
This allows to take advantage of the existence of a simpler problem,
the random-mirror problem,
for which many quantities can be computed analytically,
and then translated in terms of even-visiting walks.
Numerical checks show that the analogy between both problems
indeed works at a quantitative level.
We thus obtain in particular the estimate~(\ref{ouf}) for the number
of closed even-visiting walks, and eqs.~(\ref{cumul}), (\ref{cumula})
for the cumulants of the maximum height reached by such a walk.

\section{Definitions and mapping to a disordered system}

An $n$-step walk on a graph is a sequence $v_0v_1\cdots v_n$ of vertices
of the graph such that $v_0v_1,\dots,v_{n-1}v_n$ are edges of the graph.
An even-visiting walk is such that for each vertex $v\neq v_0,v_n$ of
the graph, the number of indices $i$ such that $v=v_i$ is even.
Note that this is a non-local constraint on the walk, so that it is
not unlikely that statistical properties of even-visiting random walks
are rather different from those of standard random walks.

In this paper we shall concentrate on one-dimensional graphs, namely
the segments with vertices $0,1,\dots,N$ or the half-line with
vertices $0,1,2,\dots$
In each case, the edges connect nearby integers.

We shall count even-visiting closed walks starting
and ending at the origin (vertex~$0$).
This problem can be reformulated as a one-dimensional
disordered system~\cite{cpv,jm}.
Consider walks on a half-line with vertices numbered $0,1,2,\dots$
Each walk comes with a weight, which is computed as follows:
each traversed edge
contributes a factor $\sqrt{t}$, and each arrival at site $i$
contributes a factor $w_i$ for $i=0,1,2,\dots$
We consider for the time being that $t$ and $w_0,w_1,\dots$
are commuting indeterminates.

Let $F(t,w_0,w_1,\dots)$ be the
generating function for walks $P$ starting and ending at the origin,
each walk being counted with its weight.
Formally
\[
F(t,w_0,w_1,\dots)=\sum_P\sqrt{t}^{S(P)}w_0^{V_0(P)} w_1^{V_1(P)}\dots,
\]
where $S(P)$ is the number of steps, and $V_i(P)$ the number of arrivals
to site $i$.
The only walk with $0$ steps has weight $1$.
Any other walk can be decomposed uniquely as a succession of elementary
blocks of the following type: a step from site $0$ to site $1$, giving
weight $\sqrt{t}\,w_1$, a closed walk on the half line made of the sites
$1,2,\dots$, contributing to $F(t,w_1,w_2,\dots)$, and a step from
site $1$ to site $0$, giving weight $\sqrt{t}\,w_0$.
So
\begin{eqnarray*}
F(t,w_0,w_1,\dots)=1&+&\sqrt{t}\,w_1F(t,w_1,w_2,\dots)\sqrt{t}\,w_0\\
&+&\left(\sqrt{t}\,w_1F(t,w_1,w_2,\dots)\sqrt{t}\,w_0\right)^2+\cdots
\end{eqnarray*}
A resummation of the geometric series gives
\begin{equation}
F(t,w_0,w_1,\dots)=\frac{1}{1-tw_0w_1F(t,w_1,w_2,\dots)},
\label{FversF}
\end{equation}
and leads to a continued-fraction expansion
\beq
F(t,w_0,w_1,\dots)=\frac{1}{{\displaystyle 1-
\frac{tw_0w_1}{{\displaystyle 1-\frac{tw_1w_2}{{\displaystyle 1-\cdots}}}}}}.
\label{jcf}
\eeq

These formul{\ae} are the starting point of all further considerations.
Note that, in the expansion of $F(t,w_0,w_1,\dots)$,
the even-visiting random walks correspond exactly to monomials such
that $V_1(P),V_2(P),\dots$ are all even.
Then $V_0(P)$ as well, because $V_0(P)+V_1(P)+\cdots$ is the number of steps,
which is even for any closed walk.
So, if the weights $w_i$ are turned into
independent random variables with vanishing odd moments and all even
moments equal to $1$, the average of $F(t,w_0,w_1,\dots)$,
denoted by $\overline{F(t,w_0,w_1,\dots)}$, is the generating
function for even-visiting walks on the half-line.
In the same way, for finite segments,
$w_0,\dots,w_N$ are chosen as above, and $w_{N+1}=0$.

Now, a random variable with vanishing odd moments and all even
moments equal to $1$ is simply a random sign:
it takes values $\pm1$ with probability $1/2$.
The above formulation can therefore be further simplified:
if $w_0,w_1,\dots w_N$ are independent random signs,
then so are $\varepsilon_N=w_0w_1$, $\varepsilon_{N-1}=w_1w_2$,
$\dots$, $\varepsilon_1=w_{N-1}w_N$.

The generating series of even-visiting walks on the segment $0,\dots,N$
reads therefore
\beq
F_N(t,\varepsilon_1,\dots,\varepsilon_N)
=\frac{1}{{\displaystyle 1-\frac{t\varepsilon_N}{{\displaystyle
1-\frac{t\varepsilon_{N-1}}{{\displaystyle 1-\frac{\cdots}{{\displaystyle
1-t\varepsilon_1}}}}}}}}.
\label{jcfrac}
\eeq
It is clear from this expression that averages
over the quenched disorder represented by the $\varepsilon_i$'s
will be even functions of $t$.
As each power of $t$ counts for two steps, the length (number of steps)
of closed even-visiting walks is a multiple of four.

Eq.~(\ref{jcfrac}) implies the recursion formula
\beq
F_N(t,\varepsilon_1,\dots,\varepsilon_N)
=\frac{1}{1-t\varepsilon_NF_{N-1}(t,\varepsilon_1,\dots,\varepsilon_{N-1})}.
\label{jrec}
\eeq

If we write $F_N=\Psi_N/\Psi_{N+1}$, with $\Psi_0=\Psi_1=1$, then we have
\beq
\Psi_{N+1}-\Psi_{N}+t\varepsilon_N\Psi_{N-1}=0.
\label{jand}
\eeq
This three-term linear recursion relation
defines a non-Hermitian one-dimensional Anderson model
with off-diagonal binary disorder.
In this context, $t$ can be interpreted as the strength of the disorder.
The variables~$F_N$, which obey the recursion relation~(\ref{jrec}),
are the associated Riccati variables~\cite{jm,ds}.

It is clear that the $t$-expansions of
$F_N(t,\varepsilon_1,\dots,\varepsilon_N)$ and
$F_{N+1}(t,\varepsilon_0,\varepsilon_1,\dots,\varepsilon_N)$
coincide up to order $N$ (a closed walk of $2N$ steps cannot visit
vertices higher than~$N$), so all formul{\ae} have a well-defined
$N\to\infty$ limit.

The $t$-expansion of $F_N$ starts with $1$,
so its power $F_N^{\alpha}$ is well-defined as a formal power series in $t$,
for $\alpha$ an arbitrary complex number, or even an indeterminate.
Surely, only when $\alpha=1$ does this quantity have a simple interpretation
as a counting problem.
Nevertheless, considering arbitrary $\alpha$ also has its own interest
(see section~8).
Anyway, we set
\[
f_N(t,\alpha)=\overline{F_N^{\alpha}(t)},\qquad
f(t,\alpha)=\lim_{N\to\infty}f_N(t,\alpha).
\]

\section{Weak-disorder expansion}

If we formally expand $F_N^{\alpha}(t)$ from eq.~(\ref{jrec})
in powers of $\varepsilon_N$, we find
\[
F_N^{\alpha}(t)=1+\sum_{m\geq1}\varepsilon_N^m F_{N-1}^m
t^m\frac{\Gamma(\alpha+m)}{\Gamma(\alpha)m!},
\]
so that after averaging (and with $m=2n$)
\[
f_N(t,\alpha)=1+\sum_{n\geq1} f_{N-1}(t,2n)t^{2n}
\frac{\Gamma(\alpha+2n)}{\Gamma(\alpha)(2n)!},
\]
and for $N\to\infty$
\begin{equation}
f(t,\alpha)=1+\sum_{n\geq1} f(t,2n)t^{2n}
\frac{\Gamma(\alpha+2n)}{\Gamma(\alpha)(2n)!}.
\label{original}
\end{equation}

Let $\N_k$ (resp.~$\N_{N,k}$) denote the number of
even-visiting closed walks of $4k$ steps starting at the origin on the
half-line (resp.~on the segment $[0,N]$),
and let us introduce the $t$-expansions
\beq
f_N(t,\alpha)=\sum_{n\ge0} f_{N,2n}(\alpha)t^{2n},
\qquad f(t,\alpha)=\sum_{n\ge0} f_{2n}(\alpha)t^{2n}.
\label{jexpan}
\eeq
Eq.~(\ref{original}) yields the following recursion relation:
\[
f_{N,2m}(\alpha)=\sum_{n=1}^m
f_{N-1,2m-2n}(2n)\frac{\Gamma(\alpha+2n)}{\Gamma(\alpha)(2n)!},
\]
with initial conditions $f_{N,0}(\alpha)=1$ for $N\geq0$.
The above formula is recursive both in $N$ and $m$,
and it holds for $N\ge1$, $m\ge1$.
All the above quantities are non-decreasing functions of $N$.
Letting $N\to\infty$, we obtain
\begin{equation}
\label{eq:f_m(alpha)}
f_{2m}(\alpha)=\sum_{n=1}^m
f_{2m-2n}(2n)\frac{\Gamma(\alpha+2n)}{\Gamma(\alpha)(2n)!}
\end{equation}
for $m\geq1$, with the initial condition $f_{0}(\alpha)=1$.

This formalism can be used to compute the number of even-visiting walks
in a very efficient way.
To do so, it is better to trade the variable $\alpha$ for integers,
to compute recursively an array $f_{2m}(2\ell)$
for $0\leq\ell\leq m\leq k-1$, using
\[
f_{2m}(2\ell)=\sum_{n=1}^mf_{2m-2n}(2n)\bin{2n+2\ell-1}{2n},
\]
and finally to obtain the number of closed even-visiting walks
with $n=4k$ steps on the half-line, starting at $0$, as
\beq
\N_k=f_{2k}(1)=\sum_{\ell=1}^kf_{2k-2\ell}(2\ell).
\label{jcaln}
\eeq
The same idea can be implemented in the case of walks on a finite segment.

With a few hours of computation on a workstation,
using the MACSYMA software, $\N_k$ has been
evaluated for $k$ up to $457$, and $\N_{N,k}$ for $N\leq k\leq125$.
Note that $\N_{N,k}=\N_k$ for $N\geq k$.
It is therefore possible to perform accurate investigations
of the number of walks and of related quantities,
such as the distribution of the maximum height (see sections~6 to 8).

It is obvious that $\N_k\leq16^k$, the total number of walks of $4k$ steps.
It is easy to generalize this (poor) bound to
see that $f_{2k}(\alpha)\leq 2^{\alpha}16^k$.
So, the radius of convergence of $f(t,\alpha)$ in $t^2$ is at least $1/16$.
In fact, for real positive $\alpha$, it is exactly $1/16$, as can be
seen from the above equations as follows.
Suppose $\ell\geq1$.
From $f(t,2\ell)=1+\sum_{m\geq1} f(t,2m)t^{2m}\bin{2\ell+2m-1}{2m}$,
we see that the $t^2$-expansion of
$\left(1-t^{2\ell}\bin{4\ell-1}{2\ell}\right)f(t,2\ell)-1=\sum_{m\neq0,\ell}
t^{2m}\bin{2\ell+2m-1}{2m}f(t,2m)$ has only non-negative coefficients.
Hence the same is true of the $t^2$-expansion of
$f(t,2\ell)-\left(1-t^{2\ell}\bin{4\ell-1}{2\ell}\right)^{-1}$.
Inserting this bound for $\ell=m$ on the
right-hand side of eq.~(\ref{original}) shows that, for $\alpha>0$,
the $t^2$-expansion of $f(t,\alpha)-1-\sum_{m\neq0}
t^{2m}\left(1-t^{2m}\bin{4m-1}{2m}\right)^{-1}\Gamma(\alpha+2m)
/(\Gamma(\alpha)(2m)!)$ has only non-negative coefficients.
Note that the radius of convergence (in powers of~$t^2$) of
$\left(1-t^{2m}\bin{4m-1}{2m}\right)^{-1}$ is very close to
$1/16$ for large $m$.
This implies that $f(t,\alpha)$ is singular at $t^2=1/16$,
giving the announced value for the radius of convergence.
We have not been able to refine the above bounds in a useful way,
starting from the weak-disorder expansion.
In the next sections we obtain much better estimates by a different method.

\section{Invariant measure and escape probability}

In order to get a feeling for the kind of singularities that appear at
$t^2=1/16$, we first evaluate a quantity apparently unrelated to the
counting of even-visiting random walks,
namely the probability for the random variable
$F(t;\varepsilon_1,\varepsilon_2,\dots)$ to be larger than $2$:
\[
P_\esc(t)={\rm Prob}(F>2).
\]
This quantity, referred to as the escape probability,
vanishes identically for $t^2\le1/16$, while it is non-zero for $t^2>1/16$.
It is quite analogous to the integrated density of states
in the Anderson model and similar disordered spectra~\cite{bl,cpv,jm,perlif}.

For the time being, consider a fixed real positive $t$, and define
the (complementary) distribution function
\[
R_N(x)={\rm Prob}(F_N>x).
\]
The recursion relation~(\ref{jrec}) implies the following relation between
the distribution functions of $F_N$ and $F_{N-1}$:
\be
2R_N(x)=R_{N-1}\left(\frac{x-1}{tx}\right)-R_{N-1}\left(\frac{1-x}{tx}\right)
+R_{N-1}\left(-\frac{1}{t}\right)-R_{N-1}\left(\frac{1}{t}\right)+2\Theta(-x),
\ee
where $\Theta$ is Heaviside's the step function:
\be
\Theta(x)=\left\{
\begin{array}{ll}0 &\mathrm{for}\; x<0,\\1 &\mathrm{for}\; x\geq0.
\end{array}\right.
\ee
For large $N$, $R_N(x)$ approaches a limiting
distribution function $R(x)$, which defines an invariant measure $\d R(x)$,
and obeys the Dyson-Schmidt equation~\cite{jm,ds}
\beq
2R(x)=R\left(\frac{x-1}{tx}\right)-R\left(\frac{1-x}{tx}\right)
+R\left(-\frac{1}{t}\right)-R\left(\frac{1}{t}\right)+2\Theta(-x).
\label{jds1}
\eeq
The fractional linear mappings involved in this equation
are the reciprocals of those involved in eq.~(\ref{jrec}).

\subsection{The case $0<t<1/4$}

In this situation, both mappings involved in eq.~(\ref{jrec})
[or in eq.~(\ref{jds1})] are hyperbolic.
Let $M(t)=(1-\sqrt{1-4t})/(2t)$ be the smallest fixed point of
the mapping $x\mapsto1/(1-tx)$, and let $m(t)=1/(1+tM(t))$.
So, $M(t)=1/(1-tM(t))$ and $-1/t<0<m(t)<1<M(t)<2<1/t$.
The mappings $x\mapsto1/(1-tx)$ and $x\mapsto1/(1+tx)$
are respectively increasing and decreasing (on their intervals of continuity).
Hence, if $x$ lies in the interval $I(t)=]m(t),M(t)[$,
then the same is true of $1/(1-tx)$ and $1/(1+tx)$.
From $F_0(t)=1$, an inductive argument shows that
$F_N(t,\varepsilon_1,\dots,\varepsilon_N)$ takes its values in $I(t)$
for all $N$.
Hence the support of the invariant measure is contained in $I(t)$.
In particular, we have $R(x)=0$ for $x\geq M(t)$
and $R(x)=1$ for $x\leq m(t)$,
so that eq.~(\ref{jds1}) can be further simplified for $x\in I(t)$ to
\beq
2R(x)=R\left(\frac{x-1}{tx}\right)-R\left(\frac{1-x}{tx}\right)+1.
\label{jds2}
\eeq
The graph of $R(x)$ is a (decreasing) devil's staircase,
with plateaus at the dyadic numbers $R=m/2^n$,
with $n=1,2,\dots$ and $1\leq m$ (odd) $\leq2^n-1$.
The support of the invariant measure $\d R(x)$,
i.e., the closure of the set of points $x$ around which $R(x)$
is not a constant, is a Cantor set of measure zero.
The heights of the plateaus do not depend on $t$,
but their size monotonically increases with $t$.
When $t\to0^+$, $R(x)$ tends to the step function $\Theta(1-x)$.
The other, more interesting limiting case $t=1/4$
is investigated in next subsection.

\subsection{The case $t=t_c=1/4$}

In this borderline case, the mapping $x\mapsto1/(1-t_cx)$ is parabolic:
it has a degenerate fixed point $x_c=2$,
which coincides with the upper bound of the interval $I(1/4)=[2/3,2]$.

The behavior of the distribution function $R(x)$ near this upper bound
will be our first example of an exponentially small, Lifshitz-like singularity.
For $x$ close to $2$, eq.~(\ref{jds2}) simplifies to
\beq
2R(x)=R\left(\frac{4(x-1)}{x}\right).
\label{jds3}
\eeq

The above equation is actually an identity for $x$ greater than $6/7$,
the image of the lower bound $2/3$ by the mapping $x\mapsto1/(1+t_cx)$.
We set
\beq
y=\frac{2}{2-x},\qquad x=2-\frac{2}{y},
\label{jy}
\eeq
so that the mapping $x\mapsto4(x-1)/x$ corresponds to $y\mapsto y-1$.
The general solution of eq.~(\ref{jds3}) therefore reads
\beq
R(x)=2^{-y}\,A(y),
\label{jrp}
\eeq
where $A(y)$ is a bounded periodic function of its argument, with unit period.
Periodic amplitudes are quite frequent in the realm
of one-dimensional disordered systems~\cite{cpv,jm,per}.
The present situation is analogous to that of the invariant measure
of the Anderson model right at a band edge~\cite{perlif}.
Figure~1 shows a plot of the periodic amplitude $A(y)$,
obtained from exactly iterating the Dyson-Schmidt equation~(\ref{jds2})
a large enough number of times,
starting from the initial data $R_0(x)=\Theta(1-x)$.

\begin{figure}[htb]
\begin{center}
\epsfig{file=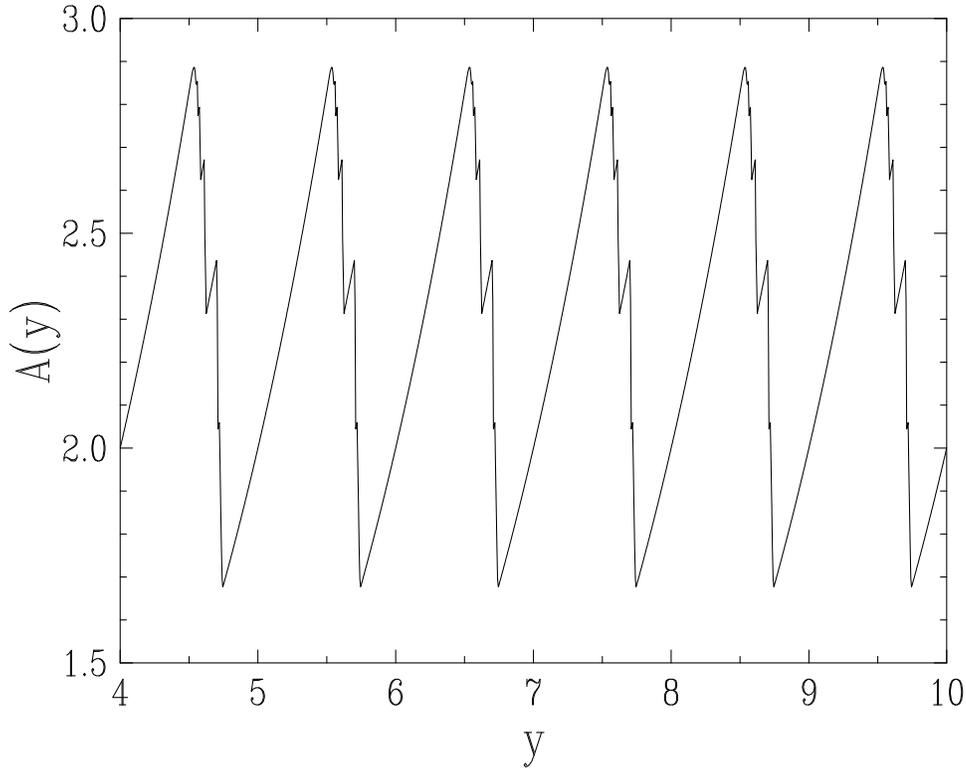,angle=90,width=.8\linewidth}
\caption{{\small
Plot of the periodic amplitude $A(y)$ entering the expression~(\ref{jrp})
of the invariant measure near its upper edge
in the critical case $(t=t_c=1/4)$.
}}
\end{center}
\end{figure}

\subsection{The case $t>1/4$}

In this situation, the mapping $x\mapsto1/(1-tx)$ is elliptic:
its two fixed points acquire an imaginary part.
With the parametrization
\beq
t=\frac{1}{4\cos^2\phi}\qquad(0<\phi<\pi/2),
\label{jcos}
\eeq
the fixed points read $x_\pm=1+\exp(\pm2\i\phi)$.
As a consequence, the support of the invariant measure is unbounded.
A fraction of the measure escapes above $2$,
hence the name `escape probability' for the quantity $P_\esc=R(2)$.

When $t-1/4\approx\phi^2/4$ is small,
the fixed points $x_\pm\approx2\pm2\i\phi$ are close to $x_c=2$,
so that the escape phenomenon takes place through a narrow channel.
It is therefore legitimate to deform the result~(\ref{jrp}) as follows.
Instead of eq.~(\ref{jy}), we set
\beq
\e^{2\i\psi}=\frac{x-x_+}{x-x_-},\qquad
x=2\cos\phi\,\frac{\sin(\psi-\phi)}{\sin\psi}\qquad(0<\psi<\pi),
\label{jpsi}
\eeq
so that the elliptic mapping $x\mapsto(x-1)/(tx)$ corresponds to
$\psi\mapsto\psi-\phi$.
The angle ratio $\psi/\phi$ is therefore the appropriate deformation
of the co-ordinate $y$ (up to an additive constant).
There is actually no such constant, as eqs.~(\ref{jy}) and~(\ref{jpsi}) yield
$\psi=y\phi+\O(\phi^3)$.
Eq.~(\ref{jrp}) therefore implies
\beq
R(x)\approx2^{-\psi/\phi}\,A(\psi/\phi),
\label{jrpsca}
\eeq
in the regime where $\phi$ is small,
while the angular variable $\psi$ is arbitrary.
In particular, $x=x_c=2$ corresponds to $\psi=\pi/2+\phi$,
hence the following prediction for the escape probability:
\beq
P_\esc\approx 2^{-\pi/(2\phi)-1}\,A\Big(\pi/(2\phi)\Big)
\approx\frac{1}{2}\,\exp\left(-\frac{\pi\ln 2}{4\sqrt{t-1/4}}\right)
\,A\left(\frac{\pi}{4\sqrt{t-1/4}}\right),
\label{escap}
\eeq
which is the result we were after.
The escape probability is therefore exponentially small in $(t-1/4)^{-1/2}$,
with an asymptotically periodic amplitude $A$,
already entering eq.~(\ref{jrp}), and shown in Figure~1.
The result~(\ref{escap}) is fully analogous to the Lifshitz tail
of the integrated density of states in the Anderson model
and similar disordered spectra~\cite{cpv,jm,perlif}.

We have not been able to obtain by direct means
estimates of the (more interesting) singularities that govern
the asymptotics of the number of even-visiting random walks.
However, the result~(\ref{escap}) for the escape probability
is quite suggestive, especially because of its
formal analogy with the Lifshitz tail for the integrated density of states
of electron or phonon spectra.
The crucial ingredient in this result
is that the transformation $x\mapsto1/(1-tx)$
turns from hyperbolic to elliptic at $t=t_c=1/4$.
Equivalently, the singularity~(\ref{escap}) is due to the occurrence
in the recursive formula~(\ref{jrec}) of regions made of a large number,
of order $\pi/(2\phi)$, of consecutive positive $\varepsilon$'s,
which are one-dimensional analogues of Lifshitz spheres.
The other transformation $x\mapsto1/(1+tx)$
has no impact on the exponentially small form of the singularity.
Its influence is only felt on the periodic modulation.

In the next section, we introduce a simpler model, the random-mirror model,
which keeps the essential feature of the original even-visiting walk problem,
i.e., the transformation $x\mapsto1/(1-tx)$, but for which
quantities of interest are directly computable by elementary means.
The random-mirror model is similar to the binary random harmonic chain
where a finite fraction of the atoms has an infinite mass,
first considered by Domb et al.~\cite{domb}.
In this limiting case, the system splits into an infinite collection
of independent finite molecules, so that the integrated density of states,
among many other quantities, can be evaluated by simple algebra.
The same simplification occurs for diffusion in the presence
of perfectly absorbing sites~\cite{jm}.
By analogy with these situations, we claim that the random-mirror model
and the even-visiting walk problem
have similar exponentially small singularities.
The simple and explicitly computable periodic amplitudes of the first model
are just replaced by (admittedly complicated) unknown periodic functions.
In the leading terms of large-order $t$-expansions,
this replacement is just responsible for an absolute prefactor.

\section{A case study: the random-mirror model}

If the distribution of the variables $\varepsilon_i$ is modified, so that they
take value $0$ or $1$ with respective probabilities $p$ or $1-p$
(before they took value $-1$ or $1$ with probability $1/2$),
the equations obtained previously for even-visiting walks are modified
in a straightforward way.
We assume $p\neq0,1$.
The physical interpretation of the present model is that
any site $i=1,2,\dots$ is totally reflecting with probability $p$,
so that the particle freely moves between the origin and the first
reflecting site, hence the name `random-mirror model'.

From a more mathematical viewpoint, this amounts to replacing the map
$x\mapsto1/(1+tx)$ by the constant map $x\mapsto1$,
without changing the second map $x\mapsto1/(1-tx)$.
We shall also consider the case of a general constant map
$x\mapsto X$ (instead of $x\mapsto1$), where $X$ is an indeterminate
(most of the time specialized to a real number $X<2$).

We shall compare in detail the random-mirror model
and the even-visiting walk model, so that it is useful to change notation:
for the random-mirror model we use $G$, $g$, and $S$,
instead of $F$, $f$, and $R$.

\subsection{Weak-disorder expansion}

Let us first fix $X=1$.
From the recursion
\[
G_N(t,\varepsilon_1,\dots,\varepsilon_N)=\frac{1}
{1-t\varepsilon_NG_{N-1}(t,\varepsilon_1,\dots,\varepsilon_{N_1})},
\]
we get
\[
g_N(t,\alpha)=\overline{G_N^{\alpha}(t)}
=1+p\sum_{n\geq1} g_{N-1}(t,n)t^{n}
\frac{\Gamma(\alpha+n)}{\Gamma(\alpha)n!},
\]
which in turn leads to a recursion relation
for the coefficients of the expansion
$g_N(t,\alpha)=\sum_n g_{N,n}(\alpha)t^{n}$, namely
\[
g_{N,m}(\alpha)=p\sum_{n=1}^m
g_{N-1,m-n}(n)\frac{\Gamma(\alpha+n)}{\Gamma(\alpha)n!},
\]
with initial conditions $g_{N,0}(\alpha)=1$ for $N\geq0$.
For $N\to\infty$, we get
\begin{equation}
\label{eq:g(t,alpha)}
g(t,\alpha)=1+p\sum_{n\geq1} g(t,n)t^{n}
\frac{\Gamma(\alpha+n)}{\Gamma(\alpha)n!},
\end{equation}
and
\begin{equation}
\label{eq:g_m(alpha)}
g_{m}(\alpha)=p\sum_{n=1}^m
g_{m-n}(n)\frac{\Gamma(\alpha+n)}{\Gamma(\alpha)n!}
\end{equation}
for $m\geq1$, with the initial condition $g_0(\alpha)=1$.
We can repeat the argument given in the previous section for
$f(t,\alpha)$ to show that for positive $\alpha$, the $t$-expansion
of $g(t,\alpha)$ has radius of convergence $1/4$.
The above formalism is easily extended to generic values of~$X$.
For example, eq.~(\ref{eq:g(t,alpha)}) becomes
\[
g(t,\alpha)=1-p+pX^{\alpha} +p\sum_{n\geq1} g(t,n)t^{n}
\frac{\Gamma(\alpha+n)}{\Gamma(\alpha)n!},
\]
while eq.~(\ref{eq:g_m(alpha)}) remains unchanged,
with the initial condition being modified to $g_0(\alpha)=1-p+pX^{\alpha}$.

\subsection{Escape probability}

We now evaluate the escape probability by more direct means
than for the case of even-visiting walks.
The random variable $G$ is equal to $Q_n$ with probability $(1-p)p^n$,
where the $Q_n$ are defined recursively as follows:
\beq
Q_{n+1}=\frac{1}{1-tQ_{n}},\qquad Q_{0}=X.
\label{jqdef}
\eeq
This definition gives a meaning to $Q_{n}$ for any integer $n$,
positive of negative,
and shows that $Q_{n}$ is a linear fractional function of $X$.
The distribution function $S(x)={\rm Prob}(G>x)$ reads explicitly
\beq
S(x)=(1-p)\sum_{n\geq0}p^n\,\Theta(Q_{n}-x).
\label{jssum}
\eeq

For the random-mirror model stricto sensu, we have $X=1$.
Suppose $t>1/4$.
With the parametrization~(\ref{jcos}),
we infer from the recursion~(\ref{jqdef}) the explicit expression
\begin{equation}
\label{eq:Qnexplicitphi}
Q_n=2\left(1-\sin\phi\,\frac{\cos(n+1)\phi}{\sin(n+2)\phi}\right).
\end{equation}
Then $Q_n\geq 2$ is equivalent to either $\sin(n+2)\phi>0$ and
$\cos(n+1)\phi\leq0$, or $\sin(n+2)\phi<0$ and $\cos(n+1)\phi\geq0$.
The smallest $n$ such that $Q_n\geq 2$ belongs to the first case, and
reads\footnote{The symbol $\lceil x\rceil$ denotes the smallest integer
larger than or equal to $x$,
while $\lfloor x\rfloor$ denotes the largest integer
smaller than or equal to $x$.
For $x$ not an integer, we have $\lceil x\rceil=\lfloor x\rfloor+1$,
and the fractional part of $x$ is defined as
$\Frac(x)=x-\lfloor x\rfloor=x-\lceil x\rceil+1$.}
$n_0=\lceil\pi/(2\phi)\rceil-1$.
When $t$ is close to $1/4$, i.e., $\phi$ is small,
many $n$'s following $n_0$, up to $n_1\approx 2n_0$, have $Q_n\geq 2$.
As a consequence, $P_\esc=S(2)$ can be estimated as
$(1-p)\sum_{n_0\leq n\leq n_1}p^n$, leading to the result
\begin{equation}
P_\esc\approx p^{\lceil\pi/(2\phi)\rceil-1},
\label{sde2}
\end{equation}
up to a correction of order $\O(p^{\pi/\phi})$.

When the initial point is not $1$, but a generic $X$,
eq.~(\ref{eq:Qnexplicitphi}) generalizes to
\begin{equation}
Q_n=2\left(1-\sin\phi\,\frac{\cos(n+1)\phi+(1-X)\cos(n-1)\phi}
{\sin(n+2)\phi+(1-X)\sin n\phi}\right).
\label{QnX}
\end{equation}
The smallest $n$ such that $Q_n\geq 2$ reads
$n_0(X)=\lceil\pi/(2\phi)-\Delta(\phi,X)\rceil$,
with
\beq
\Delta(\phi,X)=\frac{1}{\phi}\arctan\left(\frac{X}{2-X}\tan\phi\right).
\label{jdelta}
\eeq
Hence
\[
P_\esc(X)\approx p^{\lceil\pi/(2\phi)-\Delta(\phi,X)\rceil},
\]
again up to a correction of order $\O(p^{\pi/\phi})$.
For small $\phi$, this result can be recast in a form
similar to eq.~(\ref{escap}), namely
\beq
P_\esc(X)\approx p^{\pi/(2\phi)}\,A_X\Big(\pi/(2\phi)\Big),
\label{jsde2}
\eeq
where the periodic amplitude $A_X(x)$ has the explicit expression
\[
A_X(x)=p^{-\Frac(x-\Delta_0(X))+1-\Delta_0(X)},
\]
with
\[
\Delta_0(X)=\Delta(0,X)=\frac{X}{2-X}.
\]

This result shows that the random-mirror problem at $p=1/2$
and even-visiting walks have the same leading singularities
(compare eq.~(\ref{jsde2}) and eq.~(\ref{escap})):
the escape probability of
the random-mirror problem ($X=1$), of the generalized one ($X$ generic),
and of even-visiting walks only differ by their periodic modulation.
The same property is known in the case of spectra of disordered systems:
details of the distribution of random masses, random site potentials,
and so on, only enter the shape of the periodic amplitudes
of Lifshitz tails~\cite{nb,jm,perlif}.

\section{Number of even-visiting walks}

The purpose of this section is to establish the asymptotic
estimate~(\ref{ouf}) for the total number of even-visiting walks
starting and ending at the origin,
using eqs.~(\ref{jexpan}) and~(\ref{jcaln}).

The escape probability $P_\esc$ was easy to estimate
in the even-visiting walk problem,
because it is a purely singular quantity,
which vanishes as the parameter $t$ approaches $t_c=1/4$.
To the contrary, the singularities of the functions
$f(t,\alpha)$ or $g(t,\alpha)$ at $t=1/4$ are less easily grasped,
because the latter quantities also have a regular part.

We shall first consider the random-mirror model with $X=1$.
In this case, we shall see that there is a simple relation between the
discontinuity of $g(t,1)$ for $t>1/4$ and the escape probability.
A similar correspondence is expected to hold
for the discontinuity of $f(t,1)$, up to a periodic modulation.
These discontinuities provide a direct way
to estimate the large-order behavior of $t$-expansions.
The latter will lead us to estimate the number
of even-visiting random walk~(eq.~(\ref{ouf})).

\subsection{Discontinuity of $g(t,1)$}

Let us consider first the random-mirror model in the
simpler situation where $X=1$.
By analogy with eq.~(\ref{jssum}), we have the closed formula
\beq
g(t,\alpha)=(1-p)\sum_{n\geq0}p^nQ_n^{\alpha}.
\label{jg}
\eeq
Their recursive definition~(\ref{jqdef})
shows that the $Q_n(t)$ are rational functions of $t$.
Their explicit expression~(\ref{eq:Qnexplicitphi}) yields the decomposition
\begin{equation}
\label{eq:Qnexplicit}
Q_n(t)=1+\frac{4}{n+2}\sum_{m=1}^{\lfloor(n+1)/2\rfloor}
\left(\frac{1}{1-4t\cos^2\frac{\pi m}{n+2}}-1\right)\sin^2\frac{\pi m}{n+2}.
\end{equation}
It is therefore clear that all the poles of the rational function $Q_n(t)$
lie on the half-line $]1/4,+\infty[$.
We normalize the discontinuity of a real-analytic function $f(t)$
on the real axis as $D_f(t)=(\mp1/\pi)\Im f(t\pm\i0)$,
in such a way that the discontinuity of $1/t$ is exactly~$\delta(t)$.
Let $t$ be real and such that $t\geq1/4$,
and consider the integrated discontinuity
\[
D(n,t)=\int_{1/4}^t D_{Q_n}(u)\,\d u.
\]
Eq.~(\ref{eq:Qnexplicit}) yields
\beq
D(n,t)=-\frac{1}{n+2}\sum_{m=1}^{\lfloor(n+1)/2\rfloor}
\Theta\left(t-\frac{1}{4\cos^2\frac{\pi m}{n+2}}\right)\tan^2\frac{\pi m}{n+2}.
\label{jdne}
\eeq
Eq.~(\ref{jg}) implies that the integrated discontinuity of $g(t,1)$ reads
\be
D_1(t)=(1-p)\sum_{n\geq0}p^nD(n,t).
\ee
Eq.~(\ref{jdne}) leads to the following explicit result, in terms of $\phi$:
\[
D_1(\phi)=-\frac{1-p}{p^2}\sum_{n\geq 2}\frac{p^n}{n}
\sum_{m=1}^{\lfloor(n-1)/2\rfloor}
\Theta\left(\phi-\frac{\pi m}{n}\right)\tan^2\frac{\pi m}{n},
\]
which we reorganize as
\[
D_1(\phi)=-\frac{1-p}{p^2}\sum _{m\geq1}\sum_{n\geq 2m-1}\frac{p^n}{n}
\,\Theta\left(\phi-\frac{\pi m}{n}\right)\tan^2\frac{\pi m}{n}.
\]
Finally, because $0<\phi<\pi/2$, the condition
$\pi m/n\leq\phi$ is always more stringent than $n\geq 2m-1$, hence
\[
D_1(\phi)=-\frac{1-p}{p^2}\sum_{m\geq1}\sum_{n\geq\lceil\pi m/\phi\rceil}
\frac{p^n}{n}\,\tan^2\frac{\pi m}{n}.
\]
For small $\phi$, the contribution of $m=1$ is exponentially
larger than the other ones, so that the asymptotic behavior of $D_1(\phi)$ is
\begin{equation}
\label{eq:discont1}
D_1(\phi)\approx
-\frac{\phi^3}{\pi}\,p^{\lceil\pi/\phi\rceil-2}.
\end{equation}

The above formula entirely comes from the pole of $Q_n(t)$ closest
to the critical point $t_c=1/4$.
This gives the clue for treating the case of a general constant map.
In that case, $Q_n$, as given by eq.~(\ref{QnX}), has poles for
$\tan(n+1)\phi/\tan\phi=-X/(2-X)$.
Let $\phi_1(X,n)$ be the smallest solution of the latter equation.
One checks that, for large $n$,
\[
\phi_1(X,n)=\frac{\pi}{n}-\frac{2\pi}{(2-X)n^2}+\O(n^{-3}).
\]
The contributions of the closest poles for each $n$ to the total
integrated discontinuity are
\[
-(1-p)\sum_n\frac{\sin ^2\phi_1}{\cos ^3\phi_1} p^n
\frac{\cos(n+1)\phi_1 +(1-X)\cos(n-1)\phi_1}
{(n+2)\cos(n+2)\phi_1 +(1-X)n\cos n\phi_1}\,\Theta(\phi-\phi_1).
\]
This expression looks formidable,
but it simplifies drastically at small $\phi$.
From the definition of $\phi_1(X,n)$, the smallest $n$
giving a non-vanishing contribution to the above sum is such that
\[
\frac{\tan(n+1)\phi}{\tan\phi}\le-\frac{X}{2-X}<\frac{\tan n\phi}{\tan\phi},
\]
so that $n=\lceil\pi/\phi-\Delta(\phi,X)\rceil-1$,
with the definition~(\ref{jdelta}).
Hence
\[
D_X(\phi)\approx-\frac{\phi^3}{\pi}\,
p^{\lceil\pi/\phi-\Delta(\phi,X)\rceil-1},
\]
i.e.,
\beq
D_X(\phi)\approx-\frac{\phi^3}{\pi}\,p^{\pi/\phi}\,B_X(\pi/\phi),
\label{jdper}
\eeq
with
\beq
B_X(x)=p^{-\Frac(x-\Delta_0(X))-\Delta_0(X)}=\frac{A_X(x)}{p}.
\eeq
The exponentially small factor $p^{\pi/\phi}$ in the result~(\ref{jdper})
for the discontinuity of $g(t,1)$ is just the square
of the corresponding factor in the result~(\ref{jsde2})
for the escape probability.
This correspondence, which holds for any value of $X$,
is expected to hold as well for the even-visiting walk problem.
The simple relationship between the periodic amplitudes $A_X$ and $B_X$
is, however, a peculiarity of the random-mirror problem,
that we do not expect to hold in general.
We quote for further reference the constant Fourier component
of the periodic function $B_X(x)$,
\beq
B_X^{(0)}=\int_0^1 B_X(x)\,\d x=\frac{1-p}{\abs{\ln p}}\,p^{-\Delta_0(X)-1}
=\frac{1-p}{\abs{\ln p}}\,p^{-2/(2-X)}.
\label{jbx}
\eeq

\subsection{Asymptotic results}

We again consider first the generalized random-mirror model,
with arbitrary $X$.
The knowledge of the integrated discontinuity $D_X(t)$ allows to compute
the asymptotics of the $t$-expansion of $g(t,1)$.
By definition, the coefficient of order $k$ in this expansion reads
\beq
g_k(1)=\oint\frac{\d t}{2\pi\i t^{k+1}}g(t,1)=\int_{1/4}^{+\infty}
\frac{\d t}{t^{k+1}}\frac{\d D_X}{\d t}
=-\int_0^{\pi/2}(4\cos^2\phi)^{k+1}\frac{\d D_X}{\d\phi}\,\d\phi.
\label{disco}
\eeq
For large $k$, the last integral is dominated by small values of $\phi$,
where $\cos\phi$ can be expanded to quadratic order and exponentiated,
while the asymptotic result~(\ref{jdper}) holds for $D_X(\phi)$.
It turns out that only the constant Fourier component $B_X^{(0)}$
of the periodic amplitude $B_X(x)$ matters (see below eq.~(\ref{jgouf})).
We thus obtain
\beq
g_k(1)\approx 4^{k+1}\,\abs{\ln p}\,B_X^{(0)}\int_0^\infty
\phi\,\e^{-k\phi^2-\pi\abs{\ln p}/\phi}\,\d\phi.
\label{jintphi}
\eeq
This integral is then evaluated by the saddle-point approximation,
with the saddle-point value of $\phi$
being $\phi_c=(\pi\abs{\ln p}/(2k))^{1/3}$.
We are thus left with the explicit asymptotic result
\beq
g_k(1)\approx 4\,\abs{\ln p}^{4/3}\,B_X^{(0)}\left(\frac{\pi}{2k}\right)^{5/6}
\exp\left(-\frac{3}{2}\left(2\pi^2(\ln p)^2\,k\right)^{1/3}\right)4^k,
\label{jgouf}
\eeq
which only depends on $X$ through $B_X^{(0)}$~(see eq.~(\ref{jbx})).

It is now clear why only the constant component $B_X^{(0)}$ matters,
just as in trapping problems~\cite{hk,nb,jm}.
Consider the Fourier expansion $B_X(x)=\sum_k B_X^{(k)}\e^{2\pi\i kx}$.
The contribution of the Fourier component $B_X^{(k)}$
is obtained by replacing in eq.~(\ref{jgouf}) $\abs{\ln p}^{2/3}$
by $(-\ln p-2\pi\i k)^{2/3}$, a number with a strictly greater real part,
for any $k\ne0$.
The oscillations of the result~(\ref{jdper})
therefore get damped by exponentials of $k^{1/3}$
in the large-$k$ asymptotics of quantities such as $g_k(1)$.

Coming back to the even-visiting walk problem,
using the correspondence between the singularities at $t=1/4$
of $f(t,1)$ and of $g(t,1)$ with $p=1/2$,
we arrive to the following asymptotic expression
for the number $\N_k=f_{2k}(1)$ of closed even-visiting walks of $4k$ steps:
\begin{equation}
\N_k\approx B^{(0)}(2\ln 2)^{4/3}\left(\frac{\pi}{k}\right)^{5/6}
\exp\left(-\frac{3}{2}\left((2\pi\ln 2)^2\,k\right)^{1/3}\right)2^{4k}.
\label{ouf}
\end{equation}

Note that the case of even-visiting walks has the symmetry
$t\leftrightarrow-t$, so that $f(t,1)$ has two cuts,
at $]-\infty,-1/4]$ and $[1/4,+\infty[$,
which cancel each other at odd orders in $t$,
and add up constructively at even orders.

In the expression~(\ref{ouf}), $B^{(0)}$ stands for the constant component
of the periodic amplitude $B(x)$ of the even-visiting walk problem,
which cannot be predicted analytically.
The exact enumeration procedure described in section~3
has been carried out up to $k=457$, i.e., $4k=1828$ steps.
The number of these walks (exactly evaluated as an integer)
reads $\N_{457}\approx3.36829575\times10^{535}$.
Figure~2 shows a plot of the ratio $\N_k$ to its asymptotic behavior
$\N_k^{({\rm as})}$, defined as the result~(\ref{ouf}) without
its prefactor $B^{(0)}$.
The data exhibit a smooth convergence in $k^{-2/3}$.
This form of the leading finite-$k$ correction is indeed expected
for all asymptotic estimates,
as it corresponds to regular corrections to leading critical behavior,
of relative order $t-t_c$,
via the expression of the saddle-point $\phi_c$
given below eq.~(\ref{jintphi}).
We thus obtain the accurate estimate
\beq
B^{(0)}\approx1.760.
\label{jb}
\eeq

\begin{figure}[htb]
\begin{center}
\epsfig{file=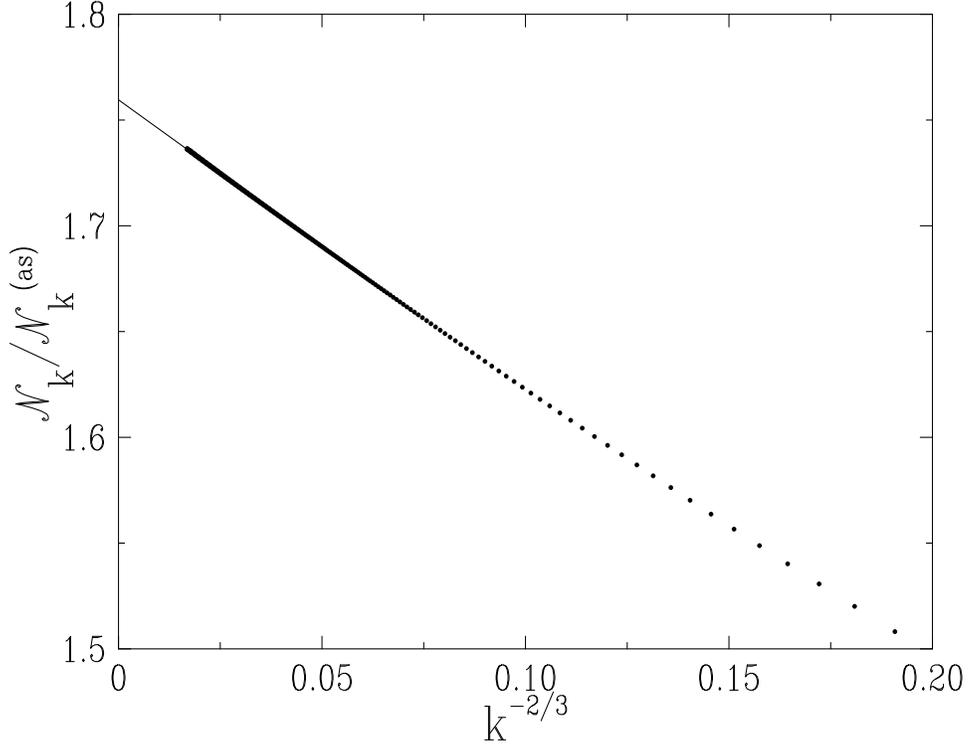,angle=90,width=.8\linewidth}
\caption{{\small
Plot of the numbers $\N_k$ of even-visiting $4k$-steps closed walks,
divided by their asymptotic form $\N_k^{({\rm as})}$
defined in the text, against $k^{-2/3}$, up to $k=457$.
Full line: least-square fit of data with $k\ge50$,
with intercept $B^{(0)}=1.760$.
}}
\end{center}
\end{figure}

\section{Distribution of the maximum height}

The purpose of this section is to investigate the distribution
of the maximum height reached by an even-visiting walk of $4k$ steps,
starting and ending at the origin.

The generating function for even-visiting walks with maximum height $M$
is the difference $f_M(t,1)-f_{M-1}(t,1)$, and its coefficient of order $2k$
is $f_{M,2k}(1)-f_{M-1,2k}(1)$.
Hence the probability that the maximum
height in an even-visiting walk of $4k$ steps be $M$~is
\beq
\Pi_{M,k}=\frac{f_{M,2k}(1)-f_{M-1,2k}(1)}{f_{2k}(1)}.
\eeq

The generating function of the cumulants of $M$ reads therefore
\beq
Z_k(z)=\sum_{j\ge1}\ll\!M^j\!\gg\frac{z^j}{j!}
=\ln\left(\sum_{M\geq0}\Pi_{M,k}\,\e^{zM}\right)
=\ln\left(\frac{1-\e^z}{f_{2k}(1)}\sum_{M\geq0}f_{M,2k}(1)\,\e^{zM}\right).
\label{jz}
\eeq

Let us again make a detour through the random-mirror model.
It is straightforward to extend eq.~(\ref{jz}) to the latter situation,
by simply replacing $f$'s by $g$'s, and $Z$ by $Y$,
even if the probabilistic interpretation of the formula thus obtained
is delicate.

The identity
\[
g_N(t,1)=(1-p)\sum_{n=0}^N p^n Q_n(t)
\]
implies
\[
Y_k(z)=\ln\frac{c_k(p\,\e^z)}{c_k(p)},
\]
with
\[
c_k(p)=\oint\frac{\d t}{2\pi\i t^{k+1}}\sum_{n\geq0} p^n Q_n(t)
=\frac{g_k(1)}{1-p}.
\]
The large-$k$ estimate~(\ref{jgouf}) for $g_k(1)$ leads to
\[
Y_k(z)\approx\frac{3}{2}\,(2\pi^2k)^{1/3}
\left((-\ln p)^{2/3}-(-\ln p-z)^{2/3}\right).
\]

Coming back to the even-visiting walk problem,
using the correspondence between both problems for $p=1/2$, we obtain
\[
Z_k(z)\approx\frac{3}{2}\,(4\pi^2k)^{1/3}
\left((\ln 2)^{2/3}-(\ln 2-z)^{2/3}\right).
\]
Hence all the cumulants of the distribution of the maximum height $M$
of $4k$-step walks are asymptotically proportional, namely
\beq
\ll\!M^j\!\gg\,\approx a_j\,k^{1/3},
\label{cumul}
\eeq
with explicit prefactors
\beq
a_j=\frac{\Gamma(j-2/3)}{\Gamma(1/3)}\,(2\pi)^{2/3}(\ln 2)^{2/3-j},
\label{cumula}
\eeq
i.e., $a_1=3.847495$, $a_2=1.850254$, $a_3=3.559136$, $a_4=11.981080$,
and so on.

As the number of steps becomes large,
the distribution of the maximum height $M$
therefore gets more and more peaked around its mean value $a_1\,k^{1/3}$.
The bulk of this distribution is a narrow Gaussian,
with standard deviation $a_2^{1/2}\,k^{1/6}$.
This result is in sharp contrast with usual random walk,
for which the largest positive extent $M$ scales as $t^{1/2}$,
and the ratio $\xi=M/t^{1/2}$ admits a non-trivial limiting
probability law~\cite{brown}.

The above analytical results have been checked against data obtained
from the exact enumeration procedure of section~3.
Figure~3 shows a plot of the first four reduced cumulants
$\ll\!M^j\!\gg\!/a_j$ $(j=1,\dots,4)$.
The clear linear behavior in $k^{1/3}$, with unit slope,
demonstrates a quantitative agreement between the exact data and the
asymptotic results~(\ref{cumul}), (\ref{cumula}).

\begin{figure}[htb]
\begin{center}
\epsfig{file=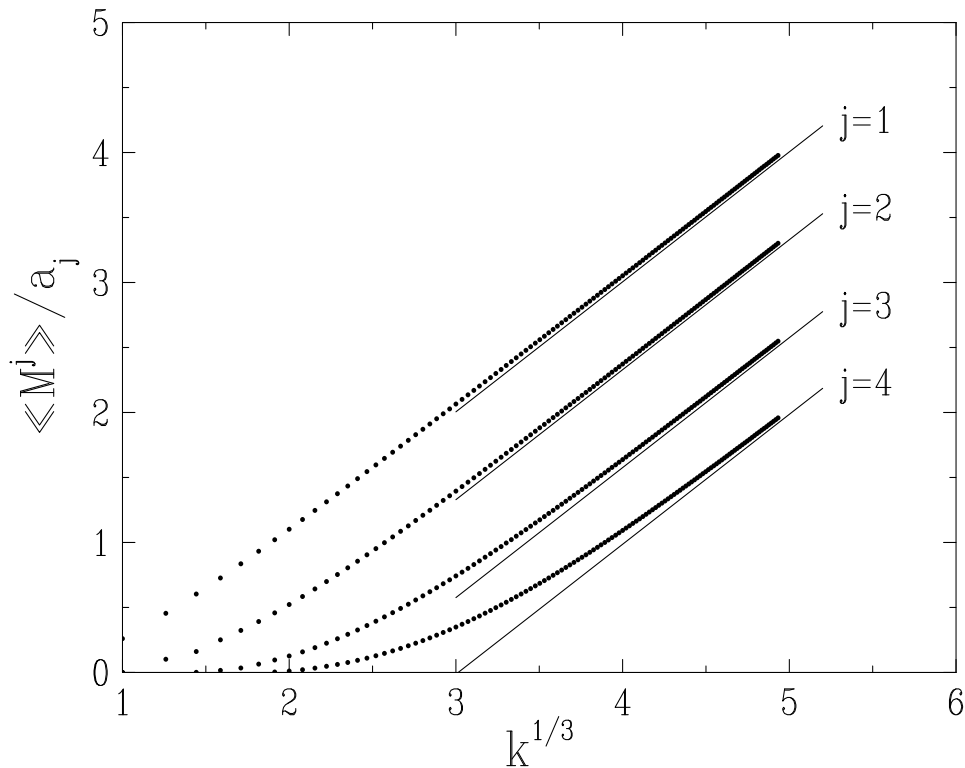,angle=90,width=.8\linewidth}
\caption{{\small
Plot of the first four reduced cumulants $\ll\!M^j\!\gg\!/a_j$
of the maximum height $M$ of $4k$-step even-visiting walks,
against $k^{1/3}$, up to $k=125$.
The full lines with unit slope demonstrate the quantitative agreement
with the analytical results~(\ref{cumul}), (\ref{cumula}).
}}
\end{center}
\end{figure}

\section{Amplitude ratios}

Up to now, we have only considered averages of Riccati variables
with respect to the quenched disorder, such as $f(t,1)$ and $g(t,1)$,
which have a direct combinatorial interpretation.
However, the even-visiting walk model and the random-mirror model
also display an interesting dependence in $\alpha$,
which will be investigated in this last section.

To start with, let us consider the random-mirror model with $X=1$,
and assume that $\alpha=\ell$ is a positive integer.
We can use eq.~(\ref{eq:Qnexplicit})
to obtain the pole structure of $Q_n^\ell(t)$,
which governs the asymptotics of its $t$-expansion.
As before, it is clear that the pole closest to the critical point ($m=1$)
gives the leading contribution.
So, we write
\begin{equation}
\label{eq:structpole}
Q_n^\ell(t)=\sum_{j=0}^\ell
\frac{c_j(n,\ell)}{\left(1-4t\cos^2\frac{\pi}{n+2}\right)^j}+\cdots,
\end{equation}
where the dots stand for a part which is regular
at $t=1/(4\cos^2\frac{\pi}{n+2})$.
It turns out that the coefficients $c_j(n,l)$ have a simple behavior
when $n$ is large.
Indeed, let us write $\phi=\pi/n-\pi\lambda/n^2+\O(1/n^3)$,
so that $t=1/(4\cos^2\frac{\pi}{n})-\pi^2\lambda/(2n^3)+\O(1/n^4)$,
and re-express $Q_{n-2}(t)$ in terms of $\lambda$.
Using eq.~(\ref{eq:Qnexplicitphi}), we can check that
$Q_{n-2}=2(1+1/\lambda)+\O(1/n)$, so that
\[
Q_{n-2}^\ell=2^\ell\left(1+\frac{1}{\lambda}\right)^\ell+\O(1/n).
\]
On the other hand, a direct substitution of the above expression for $t$
into eq.~(\ref{eq:structpole}) yields, for large $n$,
\[
Q_{n-2}^\ell\approx\sum_{j=0}^\ell
c_j(n-2,\ell)\left(\frac{2\pi ^2\lambda}{n^3}\right)^{-j}.
\]
Comparing the above two formul{\ae}, we obtain, again for large $n$,
\[
c_j(n,\ell)\approx 2^\ell\bin{\ell}{j}\left(\frac{2\pi^2}{n^3}\right)^j.
\]
Now, the $t$-expansion of eq.~(\ref{eq:structpole}) is straightforward.
A pole of order $j$ gives an extra factor $(j+k-1)!/((j-1)!k!)$
compared to a pole of order $1$.
The sum over $n$ is then performed by the saddle-point approximation,
with the saddle-point value of $n$ being
$n_c\approx(2\pi^2k/\abs{\ln p})^{1/3}$.
We are thus left with the following explicit values of the amplitude ratios:
\begin{equation}
\label{eq:amplrat}
\G(\ell)\equiv\lim_{k\to\infty}\frac{g_k(\ell)}{g_k(1)}
=2^{\ell-1}\sum_{j=1}^\ell\frac{\abs{\ln p}^{j-1}}{(j-1)!}\bin{\ell}{j}.
\end{equation}
It turns out that this result is independent of $X$, but we shall not
go into any more detail.

Quite unexpectedly, the amplitude ratios $\G(\ell)$ are also determined as
the solution of an eigenvalue problem.
The starting point is equation~(\ref{eq:g_m(alpha)}),
which we divide by $g_m(1)$.
Because the radius of convergence of the $t$-expansion
of $g(t,\alpha)$ is $1/4$, we expect that, for large $m$ and
fixed $n$, $g_{m-n}(n)\approx 4^{-n} g_m(n)$.
So, we get
\[
\G(\alpha)=p\sum_{n\geq1}\G(n) 4^{-n}
\frac{\Gamma(\alpha+n)}{\Gamma(\alpha)n!}.
\]
In particular, fixing $\alpha=\ell$, a positive integer,
\begin{equation}
\label{eq:amplratlin}
\G(\ell)=p\sum_{n\geq1}\G(n)4^{-n}\frac{(\ell+n-1)!}{(\ell-1)!n!}.
\end{equation}
This equation implies that the (infinite) matrix $\M$ with positive entries
$\M_{\ell,n}=(\ell+n-1)!/(4^n(\ell-1)!n!)$ for $\ell,n\geq1$
has an eigenvector with eigenvalue $1/p$ and positive components $\G(\ell)$,
for every $p\in]0,1]$.
We have therefore an explicit example of why the Perron--Frobenius theorem
(stating that a finite positive matrix has exactly
one eigenvector with positive components)
cannot extend to infinite matrices without further hypotheses.
Note that the matrix $\M$ is highly asymmetric,
has unbounded entries, and a divergent trace.

It is interesting to compare the linear system~(\ref{eq:amplratlin})
with eq.~(\ref{eq:g(t,alpha)}) at $t=1/4$, which reads
\[
g(1/4,\alpha)=1+p\sum_{n\geq1} g(1/4,n) 4^{-n}
\frac{\Gamma(\alpha+n)}{\Gamma(\alpha)n!}.
\]
In particular, fixing $\alpha=\ell$, a non-negative integer,
\[
g(1/4,\ell)=g(1/4,0)+p\sum_{n\geq1}
g(1/4,n) 4^{-n}\frac{(\ell+n-1)!}{(\ell-1)!n!}.
\]
So, despite the fact that analyticity for $|t|<1/4$ and continuity at
$t=1/4$ fix $g(1/4,\alpha)$ unambiguously, the $\G(\ell)$ are a (positive)
ambiguity of the solution of eq.~(\ref{eq:g(t,alpha)}) restricted to $t=1/4$.
This is again incompatible with the Perron--Frobenius property.

Because of these strange properties, it is reassuring to show
that the explicit formula~(\ref{eq:amplrat}) for the $\G(\ell)$
indeed satisfies~(\ref{eq:amplratlin}), which was derived only heuristically.
The proof relies on the following identity:
\[
\frac{1}{(\ell-1)!}\sum_{j\geq0}\frac{(j+\ell)!}{j!(j+1)!}s^j=\e^s\ell!
\sum_{j=1}^\ell\frac{s^{j-1}}{(j-1)!j!(\ell-j)!},
\]
which holds for $\ell$ a positive integer,
and can be checked by an explicit expansion of its right-hand side
in powers of $s$.
We thus obtain the explicit formula
\beq
\G(\alpha)=p\,\frac{2^{\alpha -1}}{\Gamma(\alpha)}
\sum_{j\geq0}\frac{\Gamma(j+\alpha+1)}{j!(j+1)!}\abs{\ln p}^j.
\eeq
When $\alpha$ is large and positive, at fixed $p\ne1$,
the sum over $j$ can be evaluated by the saddle-point approximation,
the saddle-point value of $j$ being
$j_c\approx(\abs{\ln p}\alpha)^{1/2}+\abs{\ln p}/2$, yielding
\beq
\G(\alpha)\approx\frac{1}{4}\left(\frac{p^2\alpha}{\pi^2\abs{\ln p}^3}
\right)^{1/4}\exp\left(2(\abs{\ln p}\alpha)^{1/2}\right)\,2^\alpha.
\label{jgag}
\eeq

In the present case of the random-mirror model,
we have been lucky enough to obtain the exact
amplitude ratios~(\ref{eq:amplrat}).
Had it not been the case, eq.~(\ref{eq:amplratlin}) would
provide a convenient tool to get accurate numerical values for these ratios.
In fact, this alternative approach leads to a better understanding of
how the Perron--Frobenius theorem is bypassed.
If the matrix $\M_{\ell,n}$ is truncated to $\ell$, $n\leq\ell_{\rm max}$,
then, whatever $\ell_{\rm max}$ is,
the spectrum does not contains any eigenvalue larger than unity.
So, the right trick is to truncate $\M_{\ell,n}$
to $\ell\leq\ell_{\rm max}$, $n\leq\ell_{\rm max}+1$,
to fix the normalization $\G(1)=1$,
and to solve for $\G(2),\dots,\G(\ell_{\rm max}+1)$.

Coming back to the even-visiting walk problem,
we shall follow the latter strategy.
We expect, by analogy with the previous situation,
that $\F(\alpha)=\lim_{k\to\infty}\Big(f_{2k}(\alpha)/f_{2k}(1)\Big)$
are well-behaved limits.
Then, starting from eq.~(\ref{eq:f_m(alpha)}),
we can repeat the above argument, to show that
\be
\F(\alpha)=\sum_{n\geq1}\F(2n)16^{-n}
\frac{\Gamma(\alpha+2n)}{\Gamma(\alpha)(2n)!},
\ee
and, fixing again $\alpha=\ell$, a non-negative integer,
\beq
\F(\ell)=\sum_{n\geq1}\F(2n)16^{-n}
\frac{(\ell+2n-1)!}{(\ell-1)!(2n)!}.
\label{jfsyst}
\eeq
This result implies that eq.~(\ref{original}) that determines $f(t,\alpha)$
is ambiguous at $t^2=1/16$.
By truncating the system~(\ref{jfsyst}) as explained above,
we can obtain the amplitude ratios $\F(\ell)$ with very high accuracy,
namely $\F(2)=5.591806$, $\F(3)=22.285850$, $\F(4)=77.059126$, and so on.

These predictions have been checked against exact data for the ratios
$f_{2k}(\ell)/f_{2k}(1)$, obtained by means of the enumeration approach
of section~3.
Figure~4 demonstrates a quantitative agreement, for $\ell=2$ to 4.
The data smoothly converge to the predicted limiting values $\F(\ell)$,
with (small) $k^{-2/3}$ corrections.
This reinforces our assertion that even-visiting walks and random-mirror
models have very similar asymptotic behaviors, up to multiplicative prefactors.

\begin{figure}[htb]
\begin{center}
\epsfig{file=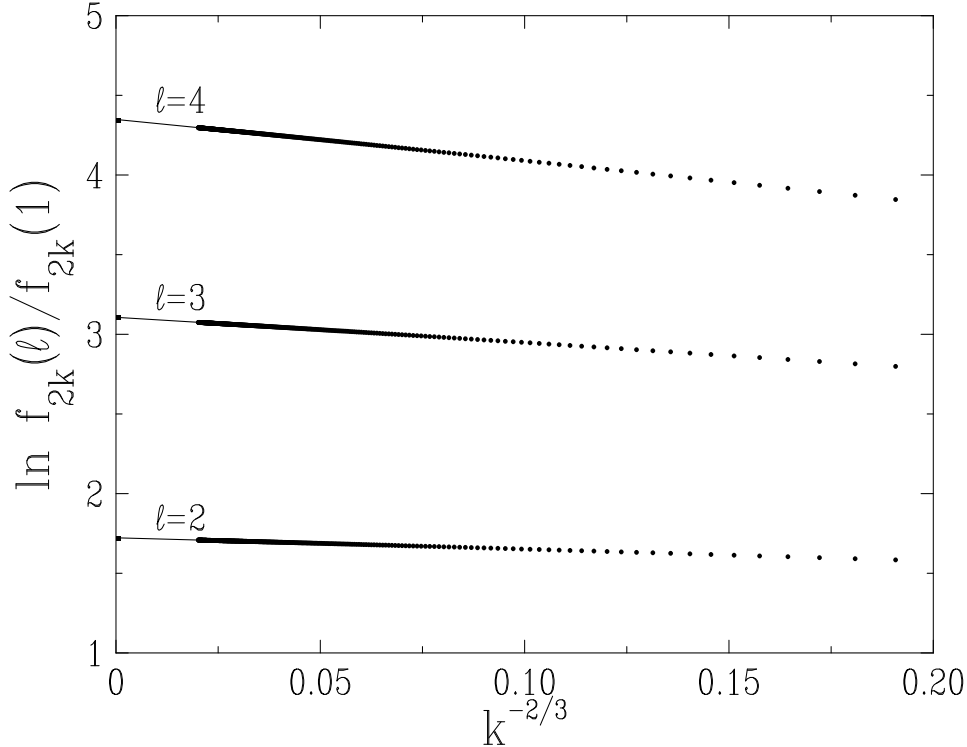,angle=90,width=.8\linewidth}
\caption{{\small
Logarithmic plot of the ratios $f_{2k}(\ell)/f_{2k}(1)$
against $k^{-2/3}$, for $\ell=2$ to 4, and $k$ up to 400.
Full lines: least-square fits of data with $k\ge50$.
The intercepts match very accurately the predicted
limiting amplitude ratios $\F(\ell)$ (symbols on vertical axis).
}}
\end{center}
\end{figure}

\section{Discussion}

In this paper we have reconsidered the problem of even-visiting random walks,
and obtained several kinds of exact or asymptotic results in one dimension.

We have mapped the even-visiting walk problem
onto a non-Hermitian Anderson model~(\ref{jand}).
The weak-disorder expansion of the latter model
provides very efficient numerical tools to enumerate
and characterize even-visiting walks.
We have thus been able to evaluate exactly, among other quantities of interest,
the total number $\N_k$ of closed even-visiting walks up to $k=457$,
i.e., $4k=1828$ steps, going thus far beyond previous works.
Indeed, our result is to be compared with the exact result~\cite{ital2}
up to $80$ steps, and with the approximate numerical simulation~\cite{dennijs}
up to around $1000$ steps.

The mapping to the non-Hermitian Anderson model~(\ref{jand})
also makes many concepts and techniques
of one-dimensional disordered systems available~\cite{cpv,jm}.
The analogy with Lifshitz tails,
which was already clearly apparent from Derrida's argument~\cite{derrida}
in any dimension,
has been corroborated at the level of analytical tools
in the one-dimensional situation.
The escape probability, investigated in section~4,
is fully analogous to the integrated density of states
in disordered spectra~\cite{bl,cpv,jm,perlif},
with an exponential behavior, modulated by an oscillatory amplitude,
which is asymptotically periodic in the relevant variable.
Periodic amplitudes are actually ubiquitous in
one-dimensional disordered systems, when the quenched disorder
has a discrete distribution~\cite{cpv,jm,per}.

In order to investigate more complex quantities,
such as the generating functions of even-visiting walks,
we make an extensive use of the random-mirror model.
The latter model is similar to the random harmonic chain
where a finite fraction of the atoms has an infinite mass,
first considered by Domb et al.~\cite{domb}:
quantities of interest are directly computable by elementary means.
The results thus obtained still hold true for the even-visiting walk problem,
the only difference being that
the simple and explicit periodic amplitudes of the first model
are replaced by unknown periodic functions.
The same phenomenon is well established
in the case of spectra of one-dimensional disordered systems:
details of the distribution of random masses, random site potentials,
and so on, only enter the periodic amplitudes
of Lifshitz tails~\cite{nb,jm,perlif}.

In the present case, just as in trapping problems~\cite{hk,nb,jm},
the quantities of most interest
just involve the constant Fourier component
of the model-specific periodic amplitudes.
For the total number of even-visiting $4k$-step walks,
the amplitude~(\ref{jb}) has been determined very accurately
by comparing the data of the exact enumeration procedure
to the analytical asymptotic estimate~(\ref{ouf}).

Our analytical investigations have also led to the prediction~(\ref{cumul})
that all the cumulants of the maximum height $M$ reached by an $n$-step
even-visiting walk scale as $n^{1/3}$, with known prefactors.
Usual random walks (for which $M/n^{1/2}$ has a non-trivial limit law)
and even-visiting ones (for which $M/n^{1/3}$ becomes more and more peaked
as $n$ gets large) are therefore very different in that respect.
We have also investigated amplitude ratios,
associated with higher moments of the Riccati variables,
which exhibit a highly non-trivial dependence in the variable $\alpha$.

Finally, the asymptotic result~(\ref{ouf}) demonstrates
that the outcome~(\ref{jn}) of Derrida's original argument
is very accurate in the one-dimensional case,
as it just misses a power of the number of steps
(in one dimension we have $\Omega=2$ and $j=\pi/2$).
The analogy with Lifshitz tails also suggests that
the situation is less under control in higher dimension:
going beyond the leading estimate~(\ref{jn}),
which involves only the volume of the Lifshitz sphere,
indeed remains a difficult open problem~\cite{cardyetal}.

\subsection*{Acknowledgments}

It is a pleasure for us to thank Bernard Derrida for interesting discussions,
and Philippe Di Francesco and Emmanuel Guitter
for making us aware of this problem,
for useful discussions at early stages of this work,
and for a critical reading of the manuscript.

\end{document}